\documentclass[aps,prd,onecolumn,groupedaddress,showpacs,nofootinbib,amssymb
]{revtex4}
\usepackage[dvips]{graphicx}
\usepackage{amssymb}
\usepackage{amsmath}
\usepackage{graphicx,,color}
\usepackage{amsfonts}
\usepackage{bm}
\usepackage{cancel}
\usepackage{comment}

\newcommand\be{\begin{equation}}
\newcommand\ee{\end{equation}}
\newcommand\nn{\nonumber \\}
\newcommand\e{\mathrm{e}}

\allowdisplaybreaks[4]

\begin{document}

\tolerance=5000

\title{Integral $F(R)$ Gravity and Saddle Point Condition as a Remedy for the $H_0$-tension}
\author{S.~Nojiri,$^{1,2}$}
\email{nojiri@gravity.phys.nagoya-u.ac.jp}
\author{S.~D.~Odintsov,$^{3,4}$}
\email{odintsov@ice.cat}
\author{V.K.~Oikonomou,$^{5}$}
\email{v.k.oikonomou1979@gmail.com;voikonomou@auth.gr}
\affiliation{$^{1)}$ Department of Physics, Nagoya University, Nagoya 464-8602, Japan \\
$^{2)}$ Kobayashi-Maskawa Institute for the Origin of Particles
and the Universe, Nagoya University, Nagoya 464-8602, Japan \\
$^{3)}$ ICREA, Passeig Luis Companys, 23, 08010 Barcelona, Spain\\
$^{4)}$ Institute of Space Sciences (IEEC-CSIC) C. Can Magrans
s/n, 08193 Barcelona, Spain\\
$^{5)}$ Department of Physics, Aristotle University of
Thessaloniki, Thessaloniki 54124, Greece}

\tolerance=5000

\begin{abstract}
In this work, we shall provide an $F(R)$ gravity theoretical
framework for solving the $H_0$-tension. Specifically, by
exploiting the $F(R)$ gravity correspondence with a scalar-tensor
theory, we shall provide a condition in which when it is
satisfied, the $H_0$-tension is alleviated. The condition that
remedies the $H_0$-tension restricts the corresponding $F(R)$
gravity, and we present in brief the theoretical features of the
constrained $F(R)$ gravity theory in both the Jordan and Einstein
frames. The condition that may remedy the $H_0$-tension is based
on the existence of a metastable de Sitter point that occurs for
redshifts near the recombination. This metastable de Sitter vacuum
restricts the functional form of the $F(R)$ gravity in the Jordan
frame. We also show that by appropriately choosing the $F(R)$
gravity, along with the theoretical solution offered for the
$H_0$-tension problem, one may also provide a unified description
of the inflationary era with the late-time accelerating era, in
terms of two extra de Sitter vacua. We propose a new approach to
$F(R)$ gravity by introducing a new class of integral $F(R)$
gravity functions, which may be wider than the usual class
expressed in terms of elementary $F(R)$ gravity functions.
Finally, the Einstein frame inflationary dynamics formalism is
briefly discussed.
\end{abstract}


\maketitle

\def\be{\begin{equation}}
\def\ee{\end{equation}}
\def\nn{\nonumber \\}
\def\e{\mathrm{e}}

\section{Introduction}

The old story of the $H_0$-tension between measurements of the
Hubble rate at high and low redshift seems to persist even to
date. Indeed, as it is pointed out in the literature, there is an
observational tension between the value of the Hubble constant
inferred from small redshifts, as in the case of observational
data coming from Type Ia supernova (SNIa) calibrated by Cepheid
observations \cite{Riess:2020fzl} and from large redshifts, as the
cosmic microwave background (CMB) observational data
\cite{Planck:2018vyg}. There exist several proposals for
eliminating or alleviating the $H_0$-tension, such as the early
dark energy proposal
\cite{Niedermann:2020dwg,Poulin:2018cxd,Karwal:2016vyq}, see also
\cite{Oikonomou:2020qah,Nojiri:2019fft}, but also radical physics
changes before $70-150\,$Myrs may also eliminate the
$H_0$-tension, see for example
\cite{Perivolaropoulos:2021jda,Perivolaropoulos:2021bds,Perivolaropoulos:2022vql,Odintsov:2022eqm}.
Also, the tension might be an artifact of the Cepheid calibration
(see \cite{Mortsell:2021nzg,Perivolaropoulos:2021jda} for
example). For a mainstream of recent articles on the
$H_0$-tension, see for example
\cite{Dai:2020rfo,He:2020zns,Nakai:2020oit,DiValentino:2020naf,Agrawal:2019dlm,Yang:2018euj,Ye:2020btb,Vagnozzi:2021tjv,
Desmond:2019ygn,OColgain:2018czj,Vagnozzi:2019ezj,
Krishnan:2020obg,Colgain:2019joh,Vagnozzi:2021gjh,Lee:2022cyh,Nojiri:2021dze,Krishnan:2021dyb,Ye:2020btb,Ye:2021iwa,Ye:2022afu}
and references therein.

In this paper, we shall provide a theoretical solution to the
$H_0$-tension, exploiting the correspondence of the Jordan frame
$F(R)$ gravity theory \cite{Nojiri:2006ri, Nojiri:2010wj,Nojiri:2017ncd, Capozziello:2011et, Faraoni:2010pgm} with a
scalar-tensor theory in the Einstein frame. Already in the context
of $F(R)$ gravity, the $H_0$-tension problem has been addressed
\cite{Odintsov:2020qzd, Wang:2020dsc}, but the result was not deemed
too successful in terms of ordinary $F(R)$ gravity models. Hence,
in this work, we shall propose a condition that may eliminate the
$H_0$-tension, considered both in the Jordan and the Einstein
frame. Specifically, $F(R)$ gravity in the Jordan frame can be
rewritten in a scalar-tensor form in the Einstein frame.
When we consider the inflationary era, the scalar field is the inflaton
and when we consider the present-day accelerating expansion, the scalar field
can be regarded as quintessence. Both the
inflationary era and the present day accelerating expansion of the
Universe, correspond to the almost flat potential of the scalar
field. However, in the case of the inflationary potential, the
potential is slightly unstable and in the case of the accelerating
expansion of the present Universe, the potential could be stable.
Although one needs to be careful about the frame because the
expansion of the Universe in the Jordan frame, which may
correspond to the physical frame observed by the observers, is
different from that in the Einstein frame, the scalar-tensor description
provides insights and an intuitive understanding of
the properties of the cosmological expansion. In the scalar-tensor
description of $F(R)$ gravity, the $H_0$-tension might be solved
by the saddle point of the Einstein frame potential of the scalar
field. We may assume a stationary point of the potential and if
the stationary point is a saddle point, the potential is unstable
for one direction but stable for another direction.
If the gradient of the potential at the vicinity of the saddle point is
positive with respect to the variation of the scalar curvature,
the curvature with a smaller value than the value of the saddle
point becomes larger. After that, it reaches the saddle point and
stays there for a limited amount of time, and after that, the
scalar curvature becomes larger again. On the other hand, if the
gradient is negative at the vicinity of the saddle point, the
curvature with a value larger than that at the saddle point becomes
smaller and reaches the saddle point. At that point, it stays there
for a while and after that, the curvature becomes smaller again.
Then near the saddle point, the potential plays the role of the
cosmological constant. Then if there is a saddle point after clear
up of the Universe with a negative gradient at the vicinity of the
saddle point, a short-lasting de Sitter era occurs, for which the
Hubble rate may correspond to the Hubble constant obtained by the
CMB observation. It is worth noting that a similar principles
work has been developed in \cite{Ye:2020btb}.

Going to more technical details, we shall consider the $F(R)$
gravity model where the function $F(R)$ is given by an integral of
some function $Q(R)$ of the scalar curvature. The function $Q(R)$
reflects the properties of the potential in the scalar-tensor
description and therefore the function $Q(R)$ directly connects
the scalar-tensor description with $F(R)$ description. Therefore
it becomes easier to construct a model for which the inflationary
era and the late-time acceleration era can be described in a
unified way \cite{Nojiri:2003ft} and simultaneously, the
$H_0$-tension problem is eliminated. With this article, we propose
a newly introduced approach to standard $F(R)$ gravity, by
introducing a new class of integral $F(R)$ gravity functions, which
belong to a wider class of models, compared with conventional
$F(R)$ gravity models.

This article is organized as follows: In section II we present the
theoretical formalism of $F(R)$ gravity in both Jordan and
Einstein frames and we introduce the saddle point condition which
may explain the $H_0$-tension. We analyze in depth the forms of
$F(R)$ gravity which may resolve and $H_0$-tension and we also
discuss the correspondence with the Einstein frame theory. In
section III, we present in brief the inflationary features of the
integral $F(R)$ gravity in the Einstein which may explain the
$H_0$-tension. Finally, the conclusions follow at the end of the paper.

\section{$F(R)$ Gravity Saddle Point Proposal and the Einstein Frame Picture}

The action of $F(R)$ gravity is given by replacing the scalar
curvature $R$ in the Einstein-Hilbert action which is,
\begin{equation}
\label{JGRG6}
S_\mathrm{EH}=\int d^4 x \sqrt{-g} \left(
\frac{R}{2\kappa^2} + \mathcal{L}_\mathrm{matter} \right)\, ,
\end{equation}
by using some appropriate function of the scalar curvature, as
follows,
\begin{equation}
\label{JGRG7}
S_{F(R)}= \int d^4 x \sqrt{-g} \left(
\frac{F(R)}{2\kappa^2} + \mathcal{L}_\mathrm{matter} \right)\, .
\end{equation}
In Eqs.~(\ref{JGRG6}) and (\ref{JGRG7}),
$\mathcal{L}_\mathrm{matter}$ is the Lagrangian density of the
perfect matter fluids. In this section, we find the saddle point
condition by using the Einstein frame picture of $F(R)$ gravity
and we propose theoretical solutions which may solve the problem
of the $H_0$-tension. We also present in brief the slow-roll
inflation formalism in the Einstein frame for completeness.

\subsection{General Properties of $F(R)$ gravity}

In this subsection, we review the general properties of $F(R)$
gravity, especially the relations between the Jordan frame and the
Einstein frame and the viability conditions that any $F(R)$
gravity must satisfy.

By varying the action (\ref{JGRG7}) with respect to the metric, we
obtain the equation of motion for $F(R)$ gravity theory as
follows,
\begin{equation}
\label{JGRG13}
G^F_{\mu\nu} \equiv \frac{1}{2}g_{\mu\nu} F - R_{\mu\nu} F_R - g_{\mu\nu} \Box F_R
+ \nabla_\mu \nabla_\nu F_R
= - \kappa^2 T_{\mu\nu}\, .
\end{equation}
Here $F_R \equiv \frac{dF(R)}{dR}$ and $T_{\mu\nu}$ is the energy
momentum tensor of the perfect matter fluids.

We can find several (in many cases exact) solutions of
Eq.~(\ref{JGRG13}). Without the presence of matter, a simple solution
is given by assuming that the Ricci tensor is covariantly
constant, that is, $R_{\mu\nu}\propto g_{\mu\nu}$. Then
Eq.~(\ref{JGRG13}) is simplified to the following algebraic equation:
\begin{equation}
\label{JGRG16}
0 = 2 F - R F_R\, .
\end{equation}
If Eq.~(\ref{JGRG16}) has a solution, then the (anti-)de Sitter
and/or Schwarzschild-(anti-)de Sitter space or the Kerr-(anti-)de
Sitter space is an exact solution in a vacuum.

We should note that we can also rewrite $F(R)$ gravity in a
scalar-tensor form. We introduce an auxiliary field $A$ and
rewrite the action (\ref{JGRG7}) of the $F(R)$ gravity in the
following form,
\begin{equation}
\label{JGRG21}
S=\frac{1}{2\kappa^2}\int d^4 x \sqrt{-g}
\left\{F_R(A)\left(R-A\right) + F(A)\right\}\, .
\end{equation}
We obtain $A=R$ by the variation of the action with respect to $A$
and by substituting the obtained equation $A=R$ into the action
(\ref{JGRG21}), we find that the action in (\ref{JGRG7}) is
reproduced. If we rescale the metric by a kind of a scale
transformation,
\begin{equation}
\label{JGRG22}
g_{\mu\nu}\to \e^\sigma g_{\mu\nu}\, ,\quad \sigma = -\ln F_R(A)\, ,
\end{equation}
we obtain the action in the Einstein frame,
\begin{align}
\label{JGRG23}
S_\mathrm{E} =& \frac{1}{2\kappa^2}\int d^4 x \sqrt{-g} \left(
R - \frac{3}{2}g^{\rho\sigma}
\partial_\rho \sigma \partial_\sigma \sigma - V(\sigma)\right) \, ,\nn
V(\sigma) =& \e^\sigma g\left(\e^{-\sigma}\right)
 - \e^{2\sigma} f\left(g\left(\e^{-\sigma}\right)\right)
=\frac{A}{F_R(A)} - \frac{F(A)}{F_R(A)^2}\, .
\end{align}
Here $g\left(\e^{-\sigma}\right)$ is given by solving the equation
$\sigma = - \ln F_R(A)$ as $A=g\left(\e^{-\sigma}\right)$. Due to
the scale transformation (\ref{JGRG22}), a coupling of the scalar
field $\sigma$ with usual matter is introduced. The mass of the
scalar field $\sigma$ is given by,
\begin{equation}
\label{JGRG24}
{m_\sigma}^2 \equiv \frac{3}{2}\frac{d^2
V(\sigma)}{d\sigma^2} =\frac{3}{2}\left\{\frac{A}{F_R(A)}
 - \frac{4F(A)}{F_R(A)^2} + \frac{1}{F_{RR}(A)}\right\}\, ,
\end{equation}
and if the mass $m_\sigma$ is not very large, there appears a
large correction to the Newton law.
Here $F_{RR}(A)\equiv \left. \frac{d^2 F(R)}{dR^2} \right|_{R=A}$.

We now also need to mention the problem of antigravity.
Eq.~(\ref{JGRG21}) indicates that the effective gravitational
coupling is given by $\kappa_\mathrm{eff}^2 =
\frac{\kappa^2}{F_R}$. Therefore when $F_R$ is negative, it is
possible to have antigravity regions. Then, we need to require
\begin{equation}
\label{FR1}
F_R > 0 \, .
\end{equation}
We should note that from the
viewpoint of the field theory, the graviton becomes a ghost in the
antigravity region.

It should be noted that the de Sitter or anti-de Sitter space
solution in Eq.~(\ref{JGRG16}) corresponds to the extremum of the
potential $V(\sigma)$. In fact, we find,
\begin{equation}
\label{FRV1}
\frac{dV(\sigma)}{dA} = \frac{F_{RR}(A)}{F_R(A)^3} \left( - AF_R(A)
+ 2F(A) \right)\, .
\end{equation}
Therefore, if Eq.~(\ref{JGRG16}) is satisfied, the scalar field
$\sigma$ should be on the local maximum or local minimum of the
potential and $\sigma$ can be a constant. When the condition (\ref
{JGRG16}) is satisfied, the mass given by (\ref{JGRG24}) has the
following form,
\begin{equation}
\label{FRV3}
{m_\sigma}^2 =
\frac{3}{2 F_R(A)} \left( - A + \frac{F_R(A)}{F_{RR}(A)} \right)\, .
\end{equation}
Therefore, in the case that the condition (\ref{FR1}) for avoiding
the antigravity holds true, the mass squared ${m_\sigma}^2$ is
positive, showing that the scalar field is on the local minimum if
\begin{equation}
\label{FRV4}
 - A + \frac{F_R(A)}{F_{RR}(A)} > 0\, .
\end{equation}
On the other hand, the scalar field is on the local maximum of
the potential if
\begin{equation}
\label{FRV5}
 - A + \frac{F_R(A)}{F_{RR}(A)} < 0\, .
\end{equation}
In this case, the mass squared ${m_\sigma}^2$ is negative. The
condition (\ref{FRV4}) is nothing but the stability condition of
the de Sitter space. Then the inflation may correspond to the
unstable de Sitter space but the late-time accelerating expansion
may correspond to the stable de Sitter space.

\begin{figure}

\begin{center}

\unitlength 0.7mm


\begin{picture}(200,60)

\put(5,10){\vector(1,0){40}}
\put(10,5){\vector(0,1){40}}

\put(55,10){\vector(1,0){40}}
\put(60,5){\vector(0,1){40}}

\put(105,10){\vector(1,0){40}}
\put(110,5){\vector(0,1){40}}

\put(155,10){\vector(1,0){40}}
\put(160,5){\vector(0,1){40}}

\put(47,10){\makebox(0,0){$A$}}
\put(10,47){\makebox(0,0){$V$}}

\put(97,10){\makebox(0,0){$A$}}
\put(60,47){\makebox(0,0){$V$}}

\put(147,10){\makebox(0,0){$R$}}
\put(110,47){\makebox(0,0){$V$}}

\put(197,10){\makebox(0,0){$R$}}
\put(160,47){\makebox(0,0){$V$}}

\put(25,5){\makebox(0,0){$R_0$}}
\put(75,5){\makebox(0,0){$R_0$}}
\put(125,5){\makebox(0,0){$R_0$}}
\put(175,5){\makebox(0,0){$R_0$}}

\put(5,55){\makebox(0,0){(a)}}
\put(55,55){\makebox(0,0){(b)}}
\put(105,55){\makebox(0,0){(c)}}
\put(155,55){\makebox(0,0){(d)}}

\thicklines

\qbezier(15,45)(20,15)(25,15)
\qbezier(25,15)(35,15)(40,45)

\qbezier(65,15)(70,35)(75,35)
\qbezier(75,35)(85,35)(90,15)

\qbezier(115,35)(120,25)(125,25)
\qbezier(125,25)(135,25)(140,15)

\qbezier(165,15)(170,25)(175,25)
\qbezier(175,25)(185,25)(190,35)


\thinlines

\put(25,10){\line(0,1){5}}
\put(75,10){\line(0,1){25}}
\put(125,10){\line(0,1){15}}
\put(175,10){\line(0,1){15}}


\end{picture}

\end{center}

\caption{\label{fig1} The typical behavior of $V(A=R)$,  (a) When ${m_\sigma}^2>0$ in  (\ref{FRV3}), which corresponds to (\ref{FRV4}),
(b) When ${m_\sigma}^2<0$ in  (\ref{FRV3}), which corresponds to (\ref{FRV5}).
(c) and (d) correspond to the cases where $V(R=R_0)$ is a saddle point, and $R$ becomes larger in the case of (c) and smaller in the case of (d). }

\end{figure}
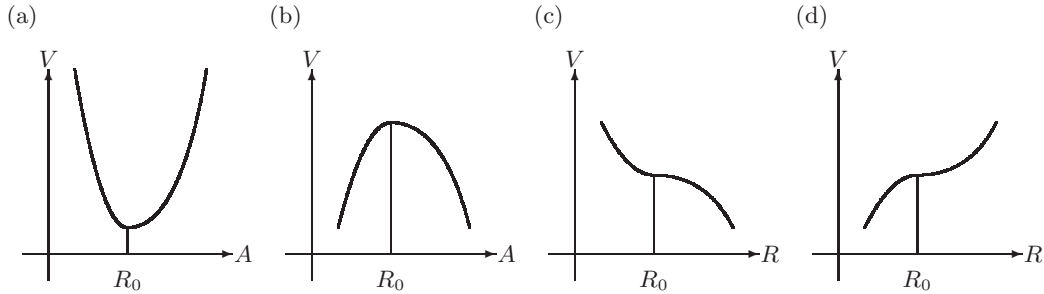

The Hubble tension might be solved by the de Sitter condition for
the saddle point. Let the solution of (\ref{JGRG16}) to be
$R=R_0$. If $R=R_0$ is a saddle point, we find $- R_0 +
\frac{F_R(R_0)}{F_{RR}(R_0)}=0$. If $\frac{dV(\sigma)}{dA}>0$ at
the vicinity of $A=R_0$, the curvature $R$ smaller than $R_0$
becomes larger and reaches $R=R_0$ and has this value for some
limited time and after that $R$ becomes larger again. On the other
hand, if $\frac{dV(\sigma)}{dA}<0$ at the vicinity of $A=R_0$, the
curvature $R$ larger than $R_0$ becomes smaller and reaches
$R=R_0$ and has this value for some limited and after that $R$
becomes smaller again. Then if there is a saddle point $R=R_0$
after clear-up of the Universe with $\frac{dV(\sigma)}{dA}<0$ at
the vicinity of $A=R_0$, a short-lasting de Sitter era is
realized, in which case, the Hubble rate corresponds to the Hubble
constant obtained by the CMB observation. In this way, the
Universe may have different Hubble rates for the CMB redshifts,
while for small redshifts the Universe has a different Hubble
rate.  In Fig.~\ref{fig1},  the typical behavior of $V(A=R)$ in
the case ${m_\sigma}^2>0$ corresponding to (\ref{FRV4}) is given
in (a) and that in the case ${m_\sigma}^2<0$ to (\ref{FRV5}) is
given in (b). With $R_0$ we denote the de Sitter saddle point
curvature. The qualitative behavior of $V(R)$ is given in detail
in Fig.~\ref{fig1}, where also the de Sitter saddle point solution
appears in all plots. Also in Fig.~\ref{fig1},  the typical
behavior of $V(A=R)$ in the case that $R$ becomes larger is given
in the subplot (c) and the case when $R$ becomes smaller is given
in the subplot (d).

\subsection{Integral $F(R)$ gravity models}

\begin{figure}

\begin{center}

\unitlength 0.7mm


\begin{picture}(200,60)

\put(5,10){\vector(1,0){185}}
\put(10,5){\vector(0,1){40}}

\put(195,10){\makebox(0,0){$R$}}
\put(10,47){\makebox(0,0){$V$}}

\put(25,5){\makebox(0,0){$R_\mathrm{late}$}}
\put(95,5){\makebox(0,0){$R_\mathrm{CMB}$}}
\put(160,5){\makebox(0,0){$R_\mathrm{inf}$}}

\thicklines

\qbezier(15,45)(20,15)(25,15)
\qbezier(25,15)(35,15)(60,25)
\qbezier(60,25)(85,35)(95,35)
\qbezier(95,35)(105,35)(120,40)
\qbezier(120,40)(135,45)(160,45)
\qbezier(160,45)(175,45)(190,35)

\thinlines

\put(25,10){\line(0,1){5}}
\put(95,10){\line(0,1){25}}
\put(160,10){\line(0,1){35}}

\end{picture}

\end{center}

\caption{\label{fig2} The quantitative behavior of $V(A=R)$ corresponding to (\ref{FRVU1}). }

\end{figure}
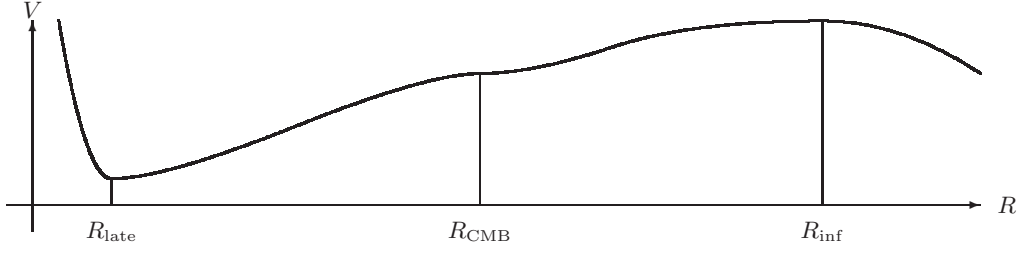

 The arguments of the previous subsection  in the last
subsection indicate that the model which solves the $H_0$-tension
problem, and simultaneously may successfully describe inflation
and the dark energy era, is given by the model which satisfies the
equation,
\begin{align}
\label{FRVU1}
\left. \frac{dV(\sigma)}{dA} \right|_{A=R}
=& \frac{d}{dR} \left( \frac{R}{F_R(R)} - \frac{F(R)}{F_R(R)^2} \right)
= \frac{F_{RR}(R)}{F_R(R)^3} \left( - AF_R(R) + 2F(R) \right) \nonumber \\
=& C(R) \left( R_\mathrm{inf} - R \right) \left( R - R_\mathrm{CMB} \right)^2
\left( R - R_\mathrm{late} \right) \, ,
\end{align}
where $C(R)$ is a positive function of $R$ or positive constant,
$R_\mathrm{inf}$ is the curvature during the inflationary epoch,
$R_\mathrm{CMB}$ is the curvature corresponding to the Hubble rate
after clear up of the Universe, and $R_\mathrm{late}$ is the
curvature in the epoch of the late-time accelerating expansion.
Therefore $R_\mathrm{inf}>R_\mathrm{CMB}>R_\mathrm{late}$. Then
for the inflationary era $R=R_\mathrm{inf}$ becomes an unstable de
Sitter era, the late-time accelerating expansion corresponding to
$R = R_\mathrm{late}$ becomes an stable de Sitter era, and the
epoch corresponding to $R = R_\mathrm{CMB}$ becomes a saddle point
with $\frac{dV(\sigma)}{dA}<0$ at the vicinity of $A=R_0$. In
Fig.~\ref{fig2},  we present the qualitative behavior of $V(A=R)$
corresponding to (\ref{FRVU1}) as a function of $R$ is given. In
the plot, several important values of the curvature are also
included, such as the values of the curvature during inflation,
dark energy era and at recombination.

We now rewrite (\ref{FRVU1}) as,
\begin{align}
\label{FRVU2}
R F_R(R) - F(R) = Q(R) F_R(R)^2 \, , \quad
Q(R) \equiv \int dR C(R) \left( R_\mathrm{inf} - R \right) \left( R - R_\mathrm{CMB} \right)^2
\left( R - R_\mathrm{late} \right) \, ,
\end{align}
which can be regarded as a differential equation for $F(R)$ with
respect to $R$. Since Eq.~(\ref{FRVU2}) can be rewritten as,
\begin{align}
\label{FRVU2B}
\frac{d}{dR} \left(\frac{F(R)}{R}\right) = Q(R) \left( \frac{F_R(R)}{R}\right)^2 \, ,
\end{align}
we can solve Eq.~(\ref{FRVU2}) as follows,
\begin{align}
\label{solFR}
F(R) = - \frac{R}{\int dR Q(R)}\, .
\end{align}
As the simplest example, we consider the case that $C(R)$ is a
positive constant $C(R)=C_0>0$. Then we find,
\begin{align}
\label{Q1}
Q(R) = C_0 &\left\{ - \frac{R^5}{5} + \frac{\left( R_\mathrm{inf} + 2 R_\mathrm{CMB} + R_\mathrm{late}\right) R^4}{4}
 - \frac{ \left( R_\mathrm{inf}R_\mathrm{late} + 2 R_\mathrm{CMB} R_\mathrm{inf}
+ 2 R_\mathrm{CMB} R_\mathrm{late} + {R_\mathrm{CMB}}^2 \right) R^3}{3} \right. \nonumber \\
& \left. \quad + \frac{ \left( 2 R_\mathrm{inf}R_\mathrm{late}
R_\mathrm{CMB} + R_\mathrm{CMB}^2 R_\mathrm{inf} + R_\mathrm{CMB}^2 R_\mathrm{late}\right) R^2}{2}
 - R_\mathrm{inf}R_\mathrm{late} R_\mathrm{CMB}^2 R + C_1 \right\} \,  ,
\end{align}
where $C_1$ is a constant of the integration. Therefore we find,
\begin{align}
\label{solFR}
F(R) = - \frac{R}{C_0} &\left\{ - \frac{R^6}{30} + \frac{\left( R_\mathrm{inf} + 2 R_\mathrm{CMB} + R_\mathrm{late}\right) R^5}{20}
 - \frac{ \left( R_\mathrm{inf}R_\mathrm{late} + 2 R_\mathrm{CMB} R_\mathrm{inf}+ 2 R_\mathrm{CMB} R_\mathrm{late}
+ {R_\mathrm{CMB}}^2 \right) R^4}{12} \right. \nonumber \\
& \left. \quad + \frac{ \left( 2 R_\mathrm{inf}R_\mathrm{late} R_\mathrm{CMB} + R_\mathrm{CMB}^2 R_\mathrm{inf}
+ R_\mathrm{CMB}^2 R_\mathrm{late}\right) R^3}{6}
 - \frac{R_\mathrm{inf}R_\mathrm{late} R_\mathrm{CMB}^2 R^2}{2} + C_1 R + C_2 \right\}^{-1} \, ,
\end{align}
were $C_2$ is a constant of the integration. For the
theory to be reduced to the Einstein-Hilbert gravity $F(R)\to R$
in the weak curvature limit, we require,
\begin{align}
\label{conss}
 - C_0 C_2 = 1 \, .
\end{align}
In Eq.~(\ref{Q1}), we considered the simplest case, where $C(R)$
is a constant. We can, however, consider the more complicated case
where $C(R)$ is a non-trivial function of $R$ although it might be
difficult to execute the integrations of $Q(R)$ in (\ref{FRVU2})
and (\ref{solFR}). If we choose $C(R)$ to be small at a stationary point corresponding to the inflationary era, the late-time
accelerating expansion, or a short inflationary era after the
clear up of the Universe, as it is clear from Eq.~(\ref{FRVU1}),
the period that the Universe remains in the vicinity the stationary point becomes larger, and on the other hand, if we
choose $C(R)$ to be large  at a stationary point, the Universe
remains at the stationary point for a short time.

\subsection{Slow-roll parameters in integral $F(R)$ gravity}

For completeness of our study, in this subsection we shall present
the formalism of inflation in the Einstein frame for integral
$F(R)$ gravity models. We shall focus on the final expressions of
the first and second slow-roll parameters in the Einstein frame.
These are defined as:
\begin{align}
\label{Eslowroll}
\epsilon_\mathrm{E} = \frac{1}{6} \left( \frac{V'(\sigma)}{V(\sigma)}\right)^2
\sim \frac{{\dot H}_\mathrm{E}}{H_\mathrm{E}^2} \, , \quad
\eta_\mathrm{E} \equiv \frac{V''(\sigma)}{3 V(\sigma)} \sim \frac{\dot H_\mathrm{E}}{H_\mathrm{E}^2}
 - \frac{1}{2} \frac{\ddot H_\mathrm{E}}{\dot H_\mathrm{E} H_\mathrm{E}}\, .
\end{align}
Here $H_\mathrm{E}$ is the Hubble rate in the Einstein frame\footnote{
We should note that the kinetic term of $\sigma$ is given by $- \frac{3}{2}g^{\rho\sigma}
\partial_\rho \sigma \partial_\sigma \sigma $ as in (\ref{JGRG23}). }.
Since $\sigma = -\ln F_R(A)$, we find $\partial_\sigma = -
\frac{F_R(A)}{F_{RR}(A)} \partial_A$. Then by using Eqs.
(\ref{JGRG23}), (\ref{JGRG24}), and (\ref{FRVU1}), we find,
\begin{align}
\label{pots}
V(\sigma) =& \frac{R}{F_R(R)} - \frac{F(R)}{F_R(R)^2}\, , \quad
V'(\sigma) = \frac{R}{F_R(R)} - \frac{2F(R) }{F_R(R)^2}  \, , \nonumber \\
V''(\sigma) =& \frac{1}{F_{RR}(R)} + \frac{R}{F_R(R)} - \frac{4F(R)}{F_R(R)^2} \, ,
\end{align}
and therefore,
\begin{align}
\label{Eslowroll2}
\epsilon_\mathrm{E} = \frac{1}{6} \left( 1 - \frac{F(R)}{RF_R(R) - F(R)} \right)^2 \, , \quad
\eta_\mathrm{E} = \frac{1}{3} \left( 1 + \frac{- 3F(R) + \frac{F_R(R)^2}{F_{RR}(R)}}{RF_R(R) - F(R)} \right) \, .
\end{align}
By using Eq.~(\ref{solFR}), we find,
\begin{align}
\label{FRR}
F(R) =& - \frac{R}{\int dR Q(R)}\, , \quad
F_R(R) = - \frac{1}{\int dR Q(R)} + \frac{RQ(R)}{\left( \int dR Q(R) \right)^2} \, , \nonumber \\
F_{RR}(R) =& \frac{2Q(R)}{\left(\int dR Q(R)\right)^2} + \frac{RQ'(R)}{\left(\int dR Q(R)\right)^2} - \frac{RQ(R)^2}{\left( \int dR Q(R) \right)^3} \, ,
\end{align}
and also,
\begin{align}
\label{pots2}
V(\sigma) =& \frac{R\left( \int dR Q(R) \right)^2}{ - \int dR Q(R) + R Q(R)}
+ \frac{ R \left( \int dR Q(R) \right)^3}{\left( - \int dR Q(R) + R Q(R) \right)^2} \, , \nonumber \\
V'(\sigma) =& \frac{R\left( \int dR Q(R) \right)^2}{ - \int dR Q(R) + R Q(R)}
+ \frac{ 2R \left( \int dR Q(R) \right)^3}{\left( - \int dR Q(R) + R Q(R) \right)^2} \, , \nonumber \\
V''(\sigma) =& \frac{\left( \int dR Q(R) \right)^3}{\left( 2Q(R) + R Q'(R) \right) \int dR Q(R) - RQ(R)^2}
+ \frac{R\left( \int dR Q(R) \right)^2}{ - \int dR Q(R) + R Q(R)}
+ \frac{4 R \left( \int dR Q(R) \right)^3}{\left( - \int dR Q(R) + R Q(R) \right)^2} \, ,
\end{align}
and in effect we have,
\begin{align}
\label{Eslowroll3}
\epsilon_\mathrm{E} =& \frac{1}{6} \left( 1 + \frac{\int dR Q(R)}{R Q(R)} \right)^2 \, ,
\quad \nonumber \\
\eta_\mathrm{E} =&  \frac{1}{3} \left( 1 + \frac{ \int dR Q(R) \left( \left( \int dR Q(R) \right)^2
+ R \left( 4Q(R) + 3RQ'(R) \right) \int dR Q(R) - 5 R^2 Q(R)^2 \right)}
{R^2 Q(R) \left( \left( 2Q(R)+RQ'(R) \right) \int dR Q(R) - 2 R Q(R)^2  \right)} \right) \, .
\end{align}
If we require $\epsilon_\mathrm{E} \sim \eta_\mathrm{E} \sim 0$,
we find,
\begin{align}
\label{Eslowroll4}
 -1 =& \frac{\int dR Q(R)}{R Q(R)} \, , \nonumber \\
 -1 =& \frac{ \int dR Q(R) \left( \left( \int dR Q(R) \right)^2
+ R \left( 4Q(R) + 3RQ'(R) \right) \int dR Q(R) - 5 R^2 Q(R)^2 \right)}
{R^2 Q(R) \left( \left( 2Q(R)+RQ'(R) \right) \int dR Q(R) - 2 R Q(R)^2  \right)} \, .
\end{align}
The first equation (\ref{Eslowroll4}) indicates,
\begin{align}
\label{Eslowroll5}
\int dR Q(R) \sim - R Q(R) \sim \frac{R_0}{R} \, ,
\end{align}
and therefore,
\begin{align}
\label{Eslowroll5}
\frac{ \int dR Q(R) \left( \left( \int dR Q(R) \right)^2
+ R \left( 4Q(R) + 3RQ'(R) \right) \int dR Q(R) - 5 R^2 Q(R)^2 \right)}
{R^2 Q(R) \left( \left( 2Q(R)+RQ'(R) \right) \int dR Q(R) - 2 R Q(R)^2  \right)} \sim -1 \, .
\end{align}
Therefore the second equation in (\ref{Eslowroll4}) is automatically satisfied if the first equation is satisfied.

\subsection{Solving $H_0$-tension by saddle point condition}

In this section, by using the outcomes of the previous
subsections,  we shall try to qualitatively describe the basic
features of integral $F(R)$ gravity models which may solve the
$H_0$-tension. Comparing Eq.~(\ref{Eslowroll5}) with
(\ref{FRVU2}), we get,
\begin{align}
\label{FRVUB1}
\left( R - R_\mathrm{late} \right) \sim \frac{2R_0}{R^3} \, ,
\end{align}
which is difficult to be realized when $R\sim R_\mathrm{inf}$ or
$R\sim R_\mathrm{late}$. Then instead of assuming that the
inflationary and the late-time accelerating expansions correspond
to stationary points, they correspond to almost flat potential
satisfying the slow-roll condition (\ref{Eslowroll5}) although
there is a saddle point corresponding to the curvature after clear
up of the Universe. As an example, instead of (\ref{FRVU1}), we
consider,
\begin{align}
\label{FRVU2B}
Q(R) = \int dR \frac{Q_0 \left( R - R_\mathrm{CMB} \right)^2}{R^5 + {Q_1}^2 R^3}\, ,
\end{align}
where $Q_0$ and $Q_1$ are positive constants. Then when $R$ is
large, we find $Q'(R) \sim \frac{Q_0}{R^3}$ and therefore $Q(R)
\sim - \frac{Q_0}{2R^2}$. On the other hand, when $R$ is small, we
find $Q'(R) \sim \frac{Q_0 {R_\mathrm{CMB}}^2}{{Q_1}^2 R^3}$ and
therefore $Q(R) \sim \frac{Q_0 {R_\mathrm{CMB}}^2}{2{Q_1}^2 R^2}$.
Therefore, both of the region where $R$ is large and the region
$R$ is small, the slow-roll conditions are satisfied. Now let us
try to realize specific cosmologies in $F(R)$ gravity, satisfying
the saddle point condition that remedies the $H_0$-tension.

The field equation corresponding to the first FRW equation in
$F(R)$ gravity is given by,
\begin{align}
\label{Hm1} 0 = - \frac{F(R)}{2} + 3 \left( H^2 + \dot H\right) F'(R)
 - 18 (\left(4 H^2 \dot H + H\ddot H\right) F''(R) + \kappa^2 \rho\ .
\end{align}
with $R=6\dot H + 12 H^2$. We now rewrite Eq.~(\ref{Hm1}) by using
a new variable (which is often called e-folding) instead of the
cosmological time $t$, $N=\ln \frac{a}{a_0}$. The variable $N$ is
related with the redshift $z$ by $\e^{-N}=\frac{a_0}{a} = 1 + z$.
Since $\frac{d}{dt} = H \frac{d}{dN}$ and therefore
$\frac{d^2}{dt^2} = H^2 \frac{d^2}{dN^2} + H \frac{dH}{dN}
\frac{d}{dN}$, one can rewrite (\ref{Hm1}) by
\begin{align}
\label{RZ4} 0 = - \frac{F(R)}{2} + 3 \left( H^2 + H H'\right) F'(R)
 - 18 (\left(4 H^3 H' + H^2 \left(H'\right)^2 + H^3 H''\right) F''(R) + \kappa^2 \rho\ .
\end{align}
Here $H'\equiv dH/dN$ and $H''\equiv d^2 H/dN^2$. If the matter
energy density $\rho$ is given by a sum of the fluid densities
with constant EoS parameter $w_i$, we find
\begin{align}
\label{RZ6} \rho=\sum_i \rho_{i0} a^{-3(1+w_i)} = \sum_i \rho_{i0}
a_0^{-3(1+w_i)} \e^{-3(1+w_i)N}\ .
\end{align}
Let assume that the Hubble rate is given in terms of $N$,
$H=H(N)$. By defining $G(N) \equiv H\left(N\right)^2$,
Eq.~(\ref{RZ4}) can be rewritten as
\begin{align}
\label{RZ11} 0 =& -9 G\left(N\left(R\right)\right)\left(4
G'\left(N\left(R\right)\right) +
G''\left(N\left(R\right)\right)\right) \frac{d^2 F(R)}{dR^2} +
\left( 3 G\left(N\left(R\right)\right) + \frac{3}{2}
G'\left(N\left(R\right)\right) \right) \frac{dF(R)}{dR} \nn & -
\frac{F(R)}{2} + \sum_i \rho_{i0} a_0^{-3(1+w_i)}
\e^{-3(1+w_i)N(R)}\ .
\end{align}
Because the scalar curvature is given by $R= 3 G'(N) + 12 G(N)$,
we have assumed that $N$ can be solved with respect to $R$,
$N=N(R)$. Hence, when we find $F(R)$ satisfying the differential
equation (\ref{RZ11}), such $F(R)$ theory admits the solution
$H=H(N)$. Hence, such $F(R)$ gravity realizes the cosmological
solution. As an example, we reconstruct the $F(R)$ gravity which
reproduces the $\Lambda$CDM-era but without the presence of any
perfect matter fluids. In the Einstein-Hilbert gravity case, the
FRW equation for the $\Lambda$CDM cosmology is given by
\begin{align}
\label{RZ13} \frac{3}{\kappa^2} H^2 = \frac{3}{\kappa^2} {H_0}^2 +
\rho_0 a^{-3} = \frac{3}{\kappa^2} {H_0}^2 + \rho_0 a_0^{-3}
\e^{-3N} \ .
\end{align}
Here $H_0$ and $\rho_0$ are constants. The first term in the
r.h.s. corresponds to the cosmological constant and the second
term to the cold dark matter (CDM). The (effective) cosmological
constant $\Lambda$ in the present Universe is given by $\Lambda =
12 {H_0}^2$. Then one gets,
\begin{align}
\label{RZ14} G(N) = {H_0}^2 + \frac{\kappa^2}{3} \rho_0 a_0^{-3}
\e^{-3N} \ ,
\end{align}
and $R = 3 G'(N) + 12 G(N) = 12 {H_0}^2 + \kappa^2\rho_0 a_0^{-3}
\e^{-3N}$, which can be solved with respect to $N$ as follows,
\begin{align}
\label{RZ16} N = - \frac{1}{3}\ln \left(\frac{ \left(R - 12
{H_0}^2\right)}{\kappa^2 \rho_0 a_0^{-3}}\right)\ .
\end{align}
Eq.~(\ref{RZ11}) takes the following form:
\begin{align}
\label{RZ17} 0=3\left(R - 9{H_0}^2\right)\left(R - 12{H_0}^2\right) \frac{d^2 F(R)}{d^2 R}
 - \left( \frac{1}{2} R - 9 {H_0}^2 \right) \frac{d F(R)}{dR} - \frac{1}{2} F(R)\ .
\end{align}
By changing the variable from $R$ to $x$ by $x=\frac{R}{3{H_0}^2}
- 3$, Eq.~(\ref{RZ17}) reduces to the hypergeometric differential
equation:
\begin{align}
\label{RZ19} 0=x(1-x)\frac{d^2 F}{dx^2} + \left(\gamma -
\left(\alpha + \beta + 1\right)x\right)\frac{dF}{dx}
 - \alpha \beta F\ .
\end{align}
Here
\begin{align}
\label{RZ20} \gamma = - \frac{1}{2}\, , \quad \alpha,\, \beta = -
\frac{1}{2}, \, \frac{1}{3}\, .
\end{align}
Solution of (\ref{RZ19}) is given by Gauss' hypergeometric
function ${_2F_1}(\alpha,\beta,\gamma;x)$:
\begin{align}
\label{RZ22} F(x) = A {_2F_1}\left(- \frac{1}{2}, \frac{1}{3}, -
\frac{1}{2}; x \right) + B x^\frac{3}{2} {_2F_1} \left(1,
\frac{11}{6}, \frac{5}{2}; x \right) \, .
\end{align}
Here $A$ and $B$ are constants. Thus, we demonstrated that
modified $F(R)$ gravity may describe the $\Lambda$CDM epoch
without the need to introduce an effective cosmological constant.
Coming back to the saddle condition that resolves the
$H_0$-tension problem, since,
\begin{align}
\label{solFR} F(R) = - \frac{R}{\int dR Q(R)}\, ,
\end{align}
and
\begin{align}
\label{GHG} {_2F_1}'(\alpha,\beta,\gamma;x)
= \frac{\alpha\beta}{\gamma} {_2F_1}(\alpha+1,\beta+1,\gamma+1;x) \, ,
\end{align}
we find,
\begin{align}
\label{GHQ} Q (R) =&\, \left[ A {_2F_1}\left(- \frac{1}{2},
\frac{1}{3}, - \frac{1}{2}; x \right) + B x^\frac{3}{2} {_2F_1}
\left(1, \frac{11}{6}, \frac{5}{2}; x \right)
 - \frac{AR}{9 {H_0}^2} {_2F_1} \left( \frac{1}{2}, \frac{4}{3},\frac{1}{2}; x \right) \right. \nonumber \\
&\,  \left. + \frac{BR}{2 {H_0}^2} x^\frac{1}{2} {_2F_1} \left(1,
\frac{11}{6}, \frac{5}{2}; x \right)
+ \frac{11B R}{45 {H_0}^2}x^\frac{3}{2} {_2F_1} \left(2, \frac{17}{6}, \frac{7}{2}; x \right) \right] \nonumber \\
&\, \times \left[ A {_2F_1}\left(- \frac{1}{2}, \frac{1}{3}, -
\frac{1}{2}; x \right) + B x^\frac{3}{2} {_2F_1} \left(1,
\frac{11}{6}, \frac{5}{2}; x \right) \right]^{-2}\, .
\end{align}
In the limit, $N\to \infty$, we find $H\to H_0$ and therefore
$R\to 12 {H_0}^2$ and $x\to 1$. Since,
\begin{align}
\label{GHG2} {_2F_1}(\alpha,\beta,\gamma;x)
=&\,\frac{\Gamma(\gamma) \Gamma(\gamma - \alpha - \beta)}
{\Gamma(\gamma - \alpha) \Gamma(\gamma - \beta)}
{_2F_1}(\alpha,\beta,\alpha + \beta + 1 -\gamma; 1-x) \nonumber \\
&\, + \frac{\Gamma(\gamma) \Gamma(- \gamma + \alpha + \beta)}
{\Gamma(\alpha) \Gamma(\beta)} \left( 1 - x \right)^{\gamma - \alpha - \beta}
{_2F_1}(\gamma - \alpha, \gamma - \beta, - \alpha - \beta + 1 + \gamma; 1-x) \, , \nonumber \\
{_2F_1}(\alpha,\beta,\gamma;0) =&\, 1 \, ,
\end{align}
we obtain,
\begin{align}
\label{GHQ3} Q (R) \to \left( A + \frac{B \Gamma \left(
\frac{5}{2} \right) \Gamma \left( \frac{1}{3} \right)}{\Gamma
\left( \frac{11}{6} \right)}\right)^{-1} \left( 4 - \frac{R}{3{H_0}^2} \right)^{- \frac{2}{3}} \frac{R}{9{H_0}^2} \, .
\end{align}
Eq.~(\ref{RZ14}) indicates that when $N$ is very large, the
contribution of the second term becomes very small and spacetime
is almost described by a de Sitter evolution. Such a situation for
which the difference from the de Sitter Universe is very small
could occur even in the inflationary era during the early
Universe.

\section{Slow-roll Parameters in Integral Form of $F(R)$ Gravity and Inflationary Dynamics Formalism}

In this section, we try to find the expressions of the slow-roll
parameters in the integral form of $F(R)$ gravity. Since
Eq.~(\ref{JGRG22}) indicates the relation between the metric
$g_{\mu\nu}$ in the Jordan frame and the metric $g_{\mathrm{E}\, \mu\nu}$
in the Einstein metric, $g_{\mu\nu}= \e^\sigma g_{\mathrm{E}\, \mu\nu}$,
if we denote the cosmological time and the scale factor in the Einstein frame
by $t_\mathrm{E}$ and $a_\mathrm{E}$, respectively, we find,
\begin{align}
\label{metric} ds^2 = - dt^2 + a(t)^2 \sum_{i=1,2,3} \left( dx^i
\right)^2 = \e^\sigma d{s}_\mathrm{E}^2 = \e^\sigma \left( -
d{t_\mathrm{E}}^2 + {a_\mathrm{E}(t)}^2 \sum_{i=1,2,3} \left(
dx_\mathrm{E}^i \right)^2 \right) \, .
\end{align}
Here we denote the quantities in the Einstein frame with the index
``$_\mathrm{E}$'' but we assume $x_\mathrm{E}^i=x^i$. Therefore,
since $dt = \e^\frac{\sigma}{2}dt_\mathrm{E}$ and
$a(t) = \e^\frac{\sigma}{2}a_\mathrm{E}(t)$, we obtain,
\begin{align}
\label{Hubble1} H_\mathrm{E} \equiv \frac{1}{a_\mathrm{E}}
\frac{d a_\mathrm{E}}{d t_\mathrm{E}}
= \frac{\e^{- \frac{\sigma}{2}}}{\e^{- \frac{\sigma}{2}}a}
\frac{d\left(\e^{- \frac{\sigma}{2}} a\right) }{dt} = \e^{- \frac{\sigma}{2}}
\left( - \frac{1}{2}\frac{d\sigma}{dt} + H \right) \, ,
\end{align}
and also,
\begin{align}
\label{Hubble2}
\frac{dH_\mathrm{E}}{d t_\mathrm{E}} =& \e^\sigma \left( \frac{1}{4}
\left( \frac{d\sigma}{dt} \right)^2 - \frac{1}{2}H\frac{d\sigma}{dt}
 - \frac{1}{2}\frac{d^2\sigma}{dt^2} + \frac{dH}{dt} \right) \, , \nonumber \\
\frac{d^2H_\mathrm{E}}{d^2 t_\mathrm{E}} =& \e^{\frac{3}{2}\sigma} \left( \frac{1}{4}
\left( \frac{d\sigma}{dt} \right)^3 - \frac{1}{2}H\left( \frac{d\sigma}{dt} \right)^2
 - \frac{1}{2}\frac{d\sigma}{dt}\frac{d^2\sigma}{dt^2} + \frac{d\sigma}{dt}\frac{dH}{dt}
+ \frac{1}{2}\frac{d\sigma}{dt}\frac{d^2\sigma}{dt^2} - \frac{1}{2} \frac{dH}{dt}
\frac{d\sigma}{dt} - \frac{1}{2}H\frac{d^2\sigma}{dt^2}
 - \frac{1}{2}\frac{d^3\sigma}{dt^3} + \frac{d^2H}{dt^2} \right) \nonumber \\
=& \e^{\frac{3}{2}\sigma} \left( \frac{1}{4} \left( \frac{d\sigma}{dt} \right)^3
 - \frac{1}{2}H\left( \frac{d\sigma}{dt} \right)^2
+ \frac{1}{2} \frac{dH}{dt} \frac{d\sigma}{dt} - \frac{1}{2}H\frac{d^2\sigma}{dt^2}
- \frac{1}{2}\frac{d^3\sigma}{dt^3} + \frac{d^2H}{dt^2} \right) \, .
\end{align}
In the Einstein frame (\ref{JGRG23}), the FRW equations have the following forms,
\begin{align}
\label{EFRW}
3 H_\mathrm{E}^2 = \frac{3}{2}\left(\frac{d \sigma}{dt_\mathrm{E}}\right)^2 + V(\sigma) \, , \quad
 - 2 \frac{d {H}_\mathrm{E}}{dt_\mathrm{E}} - 3 H_\mathrm{E}^2
= \frac{3}{2}\left(\frac{d \sigma}{dt_\mathrm{E}}\right)^2 - V(\phi) \, .
\end{align}
which gives,
\begin{align}
\label{EFRW2}
 - 2 \frac{d {H}_\mathrm{E}}{dt_\mathrm{E}} = 3\left(\frac{d \sigma}{dt_\mathrm{E}}\right)^2 \, , \quad
\frac{d {H}_\mathrm{E}}{dt_\mathrm{E}} + 3 H_\mathrm{E}^2 = V(\sigma)
\end{align}
The equation for $\sigma$ is given by,
\begin{align}
\label{feq}
0=3 \left( \frac{d^2 \sigma}{d{t_\mathrm{E}}^2} + 3 H_\mathrm{E}\frac{d \sigma}{dt_\mathrm{E}}\right) + V'(\sigma) \, .
\end{align}
By combining (\ref{EFRW2}) and (\ref{feq}), we find,
\begin{align}
\label{EFRW2BB}
\frac{d^2 {H}_\mathrm{E}}{d{t_\mathrm{E}}^2} = \frac{d \sigma}{dt_\mathrm{E}}
\left( 9 H_\mathrm{E}\frac{d \sigma}{dt_\mathrm{E}} + V'(\sigma) \right)\, .
\end{align}
By using (\ref{feq}), we also obtain,
\begin{align}
\label{feqBB}
0= 3 \left( \frac{d^3 \sigma}{d{t_\mathrm{E}}^3} + 3 \frac{d H_\mathrm{E}}{dt_\mathrm{E}}\frac{d \sigma}{dt_\mathrm{E}}
 - 9 {H_\mathrm{E}}^2\frac{d \sigma}{dt_\mathrm{E}} - H_\mathrm{E} V'(\sigma) \right)
+ V''(\sigma) \frac{d \sigma}{dt_\mathrm{E}} \, .
\end{align}
On the other hand, we find,
\begin{align}
\label{Hubble1B}
H_\mathrm{E} \equiv \frac{1}{a_\mathrm{E}}\frac{d a_\mathrm{E}}{d t_\mathrm{E}}
= \frac{\e^{- \frac{\sigma}{2}}}{\e^{- \frac{\sigma}{2}}a} \frac{d\left(\e^{- \frac{\sigma}{2}} a\right) }{dt}
= \e^{- \frac{\sigma}{2}} \left( - \frac{1}{2}\frac{d\sigma}{dt} + H \right)
= - \frac{1}{2}\frac{d\sigma}{dt_\mathrm{E}} + \e^{- \frac{\sigma}{2}} H \, .
\end{align}
and also,
\begin{align}
\label{Hubble2B}
\frac{dH_\mathrm{E}}{d t_\mathrm{E}} =&
H_\mathrm{E}\frac{d \sigma}{dt_\mathrm{E}}
+ \frac{1}{4} \left(\frac{d\sigma}{dt_\mathrm{E}} \right)^2 + \frac{1}{6}  V'(\sigma)
+ \e^{-\sigma} \frac{dH}{dt}
\, , \nonumber \\
\frac{d^2H_\mathrm{E}}{d^2 t_\mathrm{E}} =&
 - \frac{1}{2}\frac{d^3 \sigma}{{dt_\mathrm{E}}^3}
 - \frac{1}{2}\frac{d^2\sigma}{{dt_\mathrm{E}}^2} \e^{- \frac{\sigma}{2}} H
+ \frac{1}{4} \left(\frac{d\sigma}{dt_\mathrm{E}} \right)^2 \e^{- \frac{\sigma}{2}} H
 - \frac{3}{2}\frac{d\sigma}{dt_\mathrm{E}} \e^{-\sigma} \frac{dH}{dt}
+ \e^{-\frac{3}{2}\sigma} \frac{d^2H}{dt^2} \, .
\end{align}
In the slow-roll limit,
\begin{align}
\label{srll1}
\left(\frac{d \sigma}{dt_\mathrm{E}}\right)^2 \ll V(\phi) \, , \quad
\left| \frac{d^2 \sigma}{d{t_\mathrm{E}}^2} \right| \ll \left| H_\mathrm{E}\frac{d \sigma}{dt_\mathrm{E}}\right| \, ,
\end{align}
by using (\ref{EFRW}) and (\ref{feq}), we find,
\begin{align}
\label{srll2}
3 H_\mathrm{E}^2 \sim& V(\sigma) \, , \quad
 - 9 H_\mathrm{E}\frac{d \sigma}{dt_\mathrm{E}} \sim V'(\sigma) \, , \nonumber \\
\frac{d \sigma}{dt_\mathrm{E}} \sim& - \frac{V'(\sigma)}{9 \sqrt{\frac{V(\sigma)}{3}}}
= - \frac{V'(\sigma)}{3 \sqrt{3V(\sigma)}} \, , \quad
\frac{d {H}_\mathrm{E}}{dt_\mathrm{E}} = - \frac{V'(\sigma)^2 }{18 V(\sigma)} .
\end{align}
In the slow-roll limit (\ref{srll1}) and (\ref{srll2}),
Eqs.~(\ref{Hubble1B}) and (\ref{Hubble2B}) give,
\begin{align}
\label{HubbleF1}
H \sim \e^{\frac{\sigma}{2}}
\sqrt{\frac{V(\sigma)}{3}} \, , \quad \frac{dH}{dt}
\sim - \frac{1}{18} \e^\sigma V'(\sigma) \, , \quad
\frac{d^2H}{dt^2} \sim \e^{\frac{3}{2}\sigma}\frac{V'(\sigma) V''(\sigma)}
{18\sqrt{3V(\sigma)}} \, ,
\end{align}
where we have used Eq.~(\ref{srll2}). Then the slow-roll
parameters are given by,
\begin{align}
\label{slrp1}
\epsilon \equiv \frac{\dot H}{H^2} \sim - \frac{V'(\sigma)}{6V(\sigma)} \, , \quad
\eta \equiv - \frac{1}{2} \frac{\ddot H}{\dot H H} \sim \frac{V''(\sigma)}{2V(\sigma)} \, .
\end{align}
The expression of  $V(\sigma)$, $V'(\sigma)$, and $V''(\sigma)$ in
the integral form are given in (\ref{pots2}). The above relations
can be used to explicitly check the inflationary viability of an
appropriately chosen integral $F(R)$ gravity.

\section{Conclusions}

The problem of the $H_0$-tension has not been fully addressed in
the framework of vacuum $F(R)$ gravity. In this work, we
considered a theoretical solution for the $H_0$-tension, using an
$F(R)$ gravity framework. Specifically, using a saddle point
condition for the $F(R)$ gravity, and its correspondence to the
Einstein frame, we provided a theoretical framework for $F(R)$
gravity which may resolve the $H_0$-tension. This newly introduced
framework is an entirely new approach, absent in other $F(R)$
gravity frameworks and models, like for example the Starobinsky
$R^2$ model, which alone does not solve the $H_0$-tension problem.
Most of the models that can describe inflation solely, or even
inflation and dark energy in a unified way, cannot by themselves
solve the $H_0$-tension problem. One must use the integral $F(R)$
gravity formalism we introduced in this paper to produce a saddle
de Sitter point that may eventually solve the $H_0$-tension
problem. The drawback is the complicated functional form obtained,
which is not so useful for analytic calculations of the
inflationary dynamics for the model considered. The de Sitter
saddle point condition has implications in the Einstein frame, and
in the Jordan frame, the $F(R)$ gravity that achieves such a
solution has an integral form. By suitably choosing the integral
$F(R)$ gravity, apart from the $H_0$-tension, one may also unify
the inflationary era and the late-time acceleration of the
Universe. Finally, we provided the inflationary theoretical
framework which governs the integral $F(R)$ gravity. Our
theoretical solution to the $H_0$-tension problem is based on
having a saddle de Sitter point for curvature being of the order
of the CMB curvature. The formalism we introduced can be used for
producing viable inflationary $F(R)$ gravity theories, and can
also solve simultaneously in a theoretical way the $H_0$-tension
problem. Also if the formalism is viewed as a reconstruction
technique, one may manipulate appropriate functional forms of
integral $F(R)$ gravity and eventually provide a unified model for
dark energy and inflation, which simultaneously solves the
$H_0$-tension. Work is in progress toward this research line.

\section*{Acknowledgments}

This work was supported by the JSPS Grant-in-Aid for Scientific
Research (C) No. 18K03615 (S.N.) and MINECO (Spain), project
PID2019-104397GB-I00 (S.D.O). This work by S.D.O was also
partially supported by the program Unidad de Excelencia Maria de
Maeztu CEX2020-001058-M, Spain.

\end{document}